\documentclass[sigconf,natbib=true]{acmart}
\AtBeginDocument{%
  \providecommand\BibTeX{{%
    \normalfont B\kern-0.5em{\scshape i\kern-0.25em b}\kern-0.8em\TeX}}}

\usepackage[linesnumbered,ruled]
{algorithm2e}
\SetKwComment{Comment}{/* }{ */}

\usepackage{amsmath}

\copyrightyear{2023} 
\acmYear{2023} 
\setcopyright{acmlicensed}\acmConference[RecSys '23]{Seventeenth ACM Conference on Recommender Systems}{September 18--22, 2023}{Singapore, Singapore}
\acmBooktitle{Seventeenth ACM Conference on Recommender Systems (RecSys '23), September 18--22, 2023, Singapore, Singapore}
\acmPrice{15.00}
\acmDOI{10.1145/3604915.3608826}
\acmISBN{979-8-4007-0241-9/23/09}
\begin{document}

\title{Hessian-aware Quantized Node Embeddings for  Recommendation}

%\title{Compress Large-scale Node Embeddings for Bipartite Graphs}

% \author{Huiyuan Chen}
% \email{hchen@visa.com}
% \affiliation{%
%   \institution{Visa Research}
%   % \country{USA}
% }

% \author{Kaixiong Zhou}
% \email{Kaixiong.Zhou@rice.edu}
% \affiliation{%
%   \institution{Rice University}
%   %\country{USA}
% }

% \author{Kwei-Herng Lai}
% \email{khlai@rice.edu}
% \affiliation{%
%   \institution{Rice University}
%  % \country{USA}
% }

% \author{Chin-Chia Michael Yeh}
% \email{miyeh@visa.com}
% \affiliation{%
%   \institution{Visa Research}
%   %\country{USA}
% }

% \author{Yan Zheng}
% \email{yazheng@visa.com}
% \affiliation{%
%   \institution{Visa Research}
% %  \country{USA}
% }

% \author{Xia Hu}
% \email{xia.hu@rice.edu}
% \affiliation{%
%   \institution{Rice University}
% %  \country{USA}
% }

% \author{Hao Yang}
% \email{haoyang@visa.com}
% \affiliation{%
%   \institution{Visa Research}
%  % \country{USA}
% }

\author{Huiyuan Chen}
\email{hchen@visa.com}
\affiliation{%
  \institution{Visa Research}
  \city{Palo Alto}
  \state{CA}
  \country{USA}
}

\author{Kaixiong Zhou}
\author{Kwei-Herng Lai}
\email{Kaixiong.Zhou@rice.edu}
\affiliation{%
  \institution{Rice University}
  \city{Houston}
  \state{TX}
  \country{USA}
}

\author{Chin-Chia Michael Yeh}
\email{miyeh@visa.com}
\affiliation{%
  \institution{Visa Research}
  \city{Palo Alto}
  \state{CA}
  \country{USA}
}

\author{Yan Zheng}
\email{yazheng@visa.com}
\affiliation{%
  \institution{Visa Research}
  \city{Palo Alto}
  \state{CA}
  \country{USA}
}

\author{Xia Hu}
\email{xia.hu@rice.edu}
\affiliation{%
  \institution{Rice University}
  \city{Houston}
  \state{TX}
  \country{USA}
}

\author{Hao Yang}
 \email{haoyang@visa.com}
\affiliation{%
  \institution{Visa Research}
  \city{Palo Alto}
  \state{CA}
  \country{USA}
}
\renewcommand{\shortauthors}{Huiyuan Chen et al.}
\begin{abstract}
Graph Neural Networks (GNNs) have achieved  state-of-the-art  performance in recommender systems. Nevertheless, the process of  searching and ranking 
from a large item corpus    usually requires high latency, which limits the widespread deployment of GNNs in industry-scale applications. To address this issue, many  methods compress user/item representations into the binary embedding space  to reduce space requirements and accelerate inference.  Also, they use the Straight-through Estimator (STE) to prevent vanishing gradients during back-propagation.  However, the STE often causes the gradient mismatch problem, leading to sub-optimal results.

In this work, we present the Hessian-aware Quantized GNN (HQ-GNN) as an effective solution for discrete representations of users/items that enable fast retrieval. HQ-GNN is composed of two components: a GNN encoder for learning continuous node embeddings and a quantized module for compressing full-precision embeddings into low-bit ones. Consequently, HQ-GNN benefits from both lower memory requirements and faster inference speeds compared to vanilla GNNs.  To address the gradient mismatch problem in STE,  we  further consider the quantized errors and its second-order derivatives for better stability. The experimental results on several large-scale datasets show that HQ-GNN achieves a good balance between latency and performance.

\end{abstract}




\maketitle

\section{Introduction}
Recommender systems play an important role for e-commerce, such as display advertising and ranking products~\cite{huang2020embedding,chen2021tops}. Among different recommender models, Graph Neural Networks (GNNs) have achieved cutting-edge performance on top-$k$ recommendations~\cite{ying2018graph,wang2019neural,he2020lightgcn,Huang2021}. For instance,  Pinterest deploys a GNN model to train on a graph with $3$ billion nodes  and 18 billion edges, which has delivered state-of-the-art performance~\cite{ying2018graph}. Despite the superior ability of GNNs, node representations are often stored in continuous embedding space (\textsl{e.g.}, 32-bit floating point (FP32)). This often requires huge memory consumption~\cite{lian2020lightrec}. For example, the FP32 embeddings of 10 million items with a dimensional size of 256 will take up over 9.5 GB of storage space, which is hard to be deployed into devices with limited memory, especially under the federated learning settings~\cite{reisizadeh2020fedpaq,yuan2023federated}. Therefore, searching and ranking 
from a large item corpus to generate  top-$k$ recommendations  become intractable at scale due to their high latency~\cite{shi2020compositional,tan2020learning,chen2022tinykg,song2023,zhe2023}.

 Low-bit quantization~\cite{gong2019differentiable,jacob2018quantization,lee2021network,kim2021bert,cao2017hashnet} is a promising method to save the memory footprint and accelerate model inference for large-scale systems. By replacing  FP32 values with  lower precision values, e.g., 8-bit integer (INT8), quantization can shrink down the size of embeddings without modifying the original network architectures. Also,   quantized operators are widely supported by modern hardwares, which allows to deploy very large networks to resource-limited devices~\cite{jacob2018quantization,chen2022tinykg}. For example,  NVIDIA Turing GPU architecture\footnote{https://www.nvidia.com/en-us/geforce/turing/} supports the INT8 arithmetic operations.
 
 Recently, several studies have  adopted quantization in large-scale recommender systems~\cite{cao2017hashnet,tan2020learning,wu2021hashing,kang2019candidate}. However, existing  methods  suffer  from  two  drawbacks: 1) Most of them employ binary hash techniques to compress user/item embeddings into 1-bit  quantized representations. Nevertheless, recent studies show that ultra low-bit quantizations (\textsl{e.g.}, 1 or 2 bits) can be much more challenging due to their significant degradation in the accuracy~\cite{zhou2016dorefa,gong2019differentiable}; 2) They often use the Straight-through Estimator (STE)~\cite{bengio2013estimating} to avoid zero gradients  during the back-propagation. Specifically, the non-differentiable quantized function is replaced with a surrogate: the identity function~\cite{tan2020learning} or the scaled tanh function~\cite{cao2017hashnet,kang2019candidate}. However, the use of different forward and backward functions results in a gradient mismatch problem, i.e., the modified gradient is certainly not the gradient of loss function, which makes the network training unstable~\cite{yin2018understanding,chen2022learning}.
 
In this work, we propose  the Hessian-aware Quantized GNN (HQ-GNN) for effective discrete representations of users and items for fast retrieval. Specifically, HQ-GNN consists of two components: a GNN encoder for learning continuous user/item embeddings, and a quantized module for compressing the full-precision embeddings into low-bit ones. Instead of 1-bit, HQ-GNN allows arbitrary bit quantization for better trade-offs between latency and performance. To address the gradient mismatch problem, we tailor the STE by further considering the quantized errors and second-order derivatives (e.g. Hessian) for better stability and accuracy.  As such, HQ-GNN can benefit from both lower memory footprint and faster inference speed comparing to vanilla GNN.  Experimental results on several  large-scale datasets show the superiority of our HQ-GNN.

\section{Related Work}
 % In this section, we briefly review a few related lines of work. %on GNN-based recommendations and network quantizations.

\subsubsection*{\textbf{GNN-based Recommenders}}
	GNNs  have received a lot of attention  in graph domains. GNNs learn how to aggregate messages from  local neighbors using neural networks, which have been  successfully applied to user-item bipartite graphs~\cite{ying2018graph,wang2019neural,he2020lightgcn,chen2022graph, chen2022adversarial,wang2022improving}. Some representative models include  PinSage~\cite{ying2018graph}, NGCF~\cite{wang2019neural}, LightGCN~\cite{he2020lightgcn}, etc. Although GNNs have great ability of  capturing high-order collaborative signals between users and items, their node embeddings are stored in continuous space (\textsl{e.g.,} FP32), which is the major bottleneck for searching and ranking (e.g.,  high computational cost of similarity calculation between continuous embeddings). It is thus essential to improve the efficiency of generating top-$k$ recommendations at scale~\cite{shi2020compositional,tan2020learning}.
	
 \subsubsection*{\textbf{Network Quantizations}}  Quantization is a hardware-friendly approach by approximating real values with low-bit ones~\cite{gong2019differentiable,jacob2018quantization,lee2021network,kim2021bert,cao2017hashnet,jing2021meta,jiang2021xlightfm,yeh2022embedding}. Meanwhile, network inference can be performed using cheaper fixed-point multiple-accumulation  operations. As a result, quantization can reduce the storage overhead and inference latency of networks~\cite{zhou2016dorefa, gong2019differentiable,zhu2020towards,lee2021network,lian2020lightrec}. In recommender systems, HashNet~\cite{cao2017hashnet} proposes to binarize the embeddings by continuation method for multimedia retrieval. Similarly,  CIGAR~\cite{kang2019candidate}  learns binary codes to build a hash table for retrieving top-$k$ item candidates. Recently, HashGNN~\cite{tan2020learning} learns  hash functions and graph representations in an end-to-end fashion. Our HQ-GNN builds on HashGNN. Specifically, we extend 1-bit quantization of HashGNN to arbitrary-bit one, and address the gradient mismatch issue of STE, resulting in better performance.

\section{methodology}
 
\subsection{Task Description}
Generally, the input of recommender systems includes a set of users  $\mathcal{U} = \{u\}$, items $\mathcal{I}=\{i\}$, and users' implicit feedback  $\mathcal{O}^{+}=\left\{(u, i) \mid u \in \mathcal{U}, i \in \mathcal{I}, y_{ui} = 1\right\}$, where $y_{u i}=1$ indicates that user $u$ has adopted item $i$ before, $y_{u i}=0$ otherwise. One can construct a corresponding bipartite graph $\mathcal{G}=(\mathcal{V}=\mathcal{U} \cup \mathcal{I}, \mathcal{E}=\mathcal{O}^{+})$. The goal is to estimate the user preference towards unobserved items.

We next introduce our HQ-GNN that consists of two parts: a GNN encoder and a quantized module.

\subsection{GNN-based Recommenders}
Most GNNs fit under the message-passing schema~\cite{wang2019neural,he2020lightgcn}, where the representation of each node is updated    by collecting messages from its neighbors via an aggregation operation $\text{Agg}(\cdot)$ followed by an $\text{Update}(\cdot)$  operation as: 
\begin{equation}
\label{eq1}
\begin{aligned}
\mathbf{e}_{u}^{(l)}=&\text {Update}\left(\mathbf{e}_{u}^{(l-1)}, \text {Agg }(\{\mathbf{e}_{i}^{(l-1)} \mid i \in \mathcal{N}_{u}\})\right),\\
\mathbf{e}_{i}^{(l)}=&\text {Update}\left(\mathbf{e}_{i}^{(l-1)}, \text {Agg }(\{\mathbf{e}_{u}^{(l-1)} \mid u \in \mathcal{N}_{i}\})\right),
\end{aligned}
\end{equation}
where $\{\mathbf{e}^{(l)}_u, \mathbf{e}^{(l)}_i\}\in \mathbb{R}^d$ denote the embeddings of user and item in the $l$-th layer; $\mathcal{N}_u$ and $\mathcal{N}_i$ denote neighbors of user $u$ and item $i$, respectively.  By propagating $L$ layer, a pooling operator is used to obtain the final representations:
		\begin{equation}
	\label{eq2}
	\mathbf{e}_{u}=\text{Pool}(\mathbf{e}_{u}^{(0)}, \ldots, \mathbf{e}_{u}^{(L)}), \quad \mathbf{e}_{i}=\text{Pool}(\mathbf{e}_{i}^{(0)}, \ldots, \mathbf{e}_{i}^{(L)}),
	\end{equation}
	where the final representations  $\mathbf{e}_{u} \in \mathbb{R}^d$ and 	$\mathbf{e}_{i} \in \mathbb{R}^d$ can be  used  for downstream tasks. However, the full-precision embeddings, \textsl{e.g.}, FP$32$,  usually require high memory cost and power consumption to generate top-$k$ recommendations for the billion-scale graphs.

\subsection{Low-bit Quantization}
Quantization is a  hardware-friendly technique to reduce memory footprint and energy consumption~\cite{han2015deep_compression,sun2020ultra,zhu2020towards}. For a uniform $b$-bit quantization, one can clip and normalize a floating-point number $x$ into a quantization interval, parameterized by an upper $u$ and a lower $l$ bounds, as:
\begin{equation}
\label{eq3}
    x_n = \frac{\text{clip}(x, l, u) -l }{\Delta},
\end{equation}
where $x_n$ is the normalized output, $\text{clip}(x, l, u) = \min(\max(x, l),u)$, $\Delta =\frac{u-l}{2^b-1}$ is the interval length, and $b$ denotes the number of quantization levels, \textsl{e.g.}, $b=8$ for $8$-bit quantization.
During training, the clipping interval $(l, u)$ is often unknown beforehand, two strategies are commonly used to determine the upper/lower thresholds: exponential moving averages~\cite{jacob2018quantization} and treating the thresholds as learnable parameters~\cite{choi2018pact}.  The normalized output $x_n$ can be then converted to a discrete value $x_b$ using a round function with post-scaling  as~\cite{zhou2016dorefa, gong2019differentiable,zhu2020towards}:
\begin{equation}
\label{eq4}
    x_b = x_q   \cdot \Delta, \quad x_q = \text{round}(x_n),
\end{equation}
where  $\text{round}(\cdot)$  maps a full-precision value to its nearest integer. The quantized tensor $x_b$ can be then used for efficient computation by emergent  accelerators (\textsl{e.g.},  NVIDIA TensorRT) that are able to handle $\Delta$ efficiently.

By combining Eq. (\ref{eq3}) and Eq. (\ref{eq4}),  we can defined a quantization function $Q_b(\cdot)$ as: $x_b = Q_b(x)$. If the input is a vector/matrix,   $Q_b(\cdot)$ would apply to each element of the vector/matrix. To this end, we can quantize  the  GNN embeddings $\mathbf{e}_{u}$ and $\mathbf{e}_{i}$ in Eq. (\ref{eq2}) into:
\begin{equation}
	\label{eq5}
	\mathbf{q}_{u}=Q_b(\mathbf{e}_u), \quad 	\mathbf{q}_{i}=Q_b(\mathbf{e}_i),
\end{equation}
where $\{\mathbf{q}_{u}, \mathbf{q}_{i} \}\in \mathbb{R}^d$ are the $b$-bit representations of user $u$ and item $i$, respectively. Our model follows the mixed-precision quantization policy~\cite{micikevicius2018mixed}, where we only compress the \textsl{activations} of GNNs for faster inference, and leave the \textsl{weights} of GNNs at full precision. Since GNNs often contain less than three layers and have limited weights, the mixed-precision scheme  could achieve good trade-offs between performance and memory size~\cite{dong2019hawq}. The mixed-precision quantization has also become more and more common in  deep learning frameworks\footnote{https://www.tensorflow.org/guide/mixed\_precision}. 
% We are aware that some integer-only quantization schemes might  accelerate both training and inference of our model~\cite{jacob2018quantization,kim2021bert}. We leave this extension in the future.

However, the non-differentiable quantized processes are undesirable for the standard back-propagation, i.e., the quantization function is intrinsically a discontinuous step function and nearly has zero gradients, which significantly affects the training of HQ-GNN. We next present a Generalized Straight-Through Estimator to address this problem.

\subsection{Generalized Straight-Through Estimator} 
%  \begin{algorithm}
%     \small
%     \DontPrintSemicolon % Some LaTeX compilers require you to use \dontprintsemicolon instead
%     \KwIn{First-order gradient $\mathcal{G}_{\mathbf{x_q}}$.}
    
%     \KwOut{Hessian trace.}
    
%     \For{$i=1$ \KwTo $m$}
% {   
% Draw a random vector $\mathbf{v}$ from Rademacher distribution;\;
% Compute forward loss $l = \mathcal{G}^\top_\mathbf{x_q}\mathbf{v}$;\;
% Compute  $ \mathbf{H}\mathbf{v}$ by backpropagation: $ \mathbf{H}\mathbf{v} =  \frac{\partial l}{\partial \mathbf{x_q}}$ ;\;
% Compute and cache $ \mathbf{v}^\top (\mathbf{H} \mathbf{v})$;\;
% }
% Compute $\text{Tr}(\mathbf{H}) = \frac{1}{m} \sum_{i=1}^{m} ({\mathbf{v}^{(i)}}^\top \mathbf{H} \mathbf{v}^{(i)})$ for all sampled vectors;\;
%     \Return {$\text{Tr}(\mathbf{H})$}
%     \caption{Hutchinson Method for Hessian Trace}
%     \label{alg1}
% \end{algorithm}

The main challenge of training our  HQ-GNN
arises from the discretized round function in Eq. (4), where its derivative is either infinite or zero at almost everywhere. One popular family of estimators are the so-called Straight-Through Estimators (STE)~\cite{bengio2013estimating,yin2018understanding}. In STE, the forward computation of  $\text{round}(\cdot)$ is unchanged, but back-propagation is computed through a surrogate~\cite{tan2020learning,cao2017hashnet,zhou2016dorefa}: replacing $\text{round}(\cdot)$  with an identity function, \textsl{i.e.}, $\mathcal{G}_\mathbf{x_n} = \mathcal{G}_\mathbf{x_q}$ where $\mathcal{G}$ denotes the gradient operator.  However,  STE  runs the risk of convergence to poor minima and unstable training ~\cite{yin2018understanding}. For example, both values of $0.51$ and $1.49$ round to same integer $1$ with  different quantized errors.
Moreover, STE forces to update both values equally with the same gradient at integer $1$, which is likely to be biased with cumulative quantized errors.  Moreover, a small decrement (\textsl{e.g.}, $-0.2$) for value $0.51$ can largely change the quantized integer from $1$ to $0$, while a same decrement to $1.49$ cannot.

 To mitigate the impact of quantized errors, we  generalize the STE as~\cite{lee2021network}:
 \begin{equation}
 \label{eq7}
\mathcal{G}_\mathbf{x_{n}}=\mathcal{G}_\mathbf{x_{q}} \odot \left(1+\delta \cdot \text{sign}(\mathcal{G}_\mathbf{x_q})\odot(\mathbf{x_{n}}-\mathbf{x_{q}})\right),
\end{equation}
where $\odot$ denotes element-wise product; $\text{sign}(\cdot)$ is a sign function such that $\text{sign}(x) = +1$ if $x \ge0$, $-1$ otherwise; $\delta$ is the scaling factor. Eq. (\ref{eq7}) is able to scale up/down the gradient of $\mathcal{G}_\mathbf{x_{q}}$ when the $\mathbf{x_n}$ requires a larger/smaller magnitude for an update. Moreover,  Eq. (\ref{eq7}) is equivalent to vanilla STE when setting $\delta=0$. It is thus crucial to determine the scaling factor $\delta$ during training.

Inspired by   Hessian-aware quantized networks~\cite{dong2019hawq,dong2020hawq}, we use second-order information to guide the selection of $\delta$.   Let $\mathbf{\epsilon} = \mathbf{x_n}-\mathbf{x_q}$ denote the quantized error for round function, where each element of $\mathbf{\epsilon}$  is well bound by a small number, \textsl{i.e.}, $|\epsilon_i| \le \frac{0.5}{2^b-1}$, with element-wise Taylor  expansion, we have:
\begin{equation*}
\begin{aligned}
    \mathcal{G}_\mathbf{x_n} = & \mathcal{G}_\mathbf{x_q}  + \frac{\mathcal{G}_\mathbf{x_n} - \mathcal{G}_\mathbf{x_q} }{\mathbf{x_n}-\mathbf{x_q}} \odot (\mathbf{x_n}-\mathbf{x_q})\\
    = & \mathcal{G}_\mathbf{x_q}  +\frac{\mathcal{G}_{\mathbf{x_q}+\mathbf{\epsilon}} - \mathcal{G}_\mathbf{x_q} }{\mathbf{\epsilon}} \odot (\mathbf{x_n}-\mathbf{x_q}) \\
    \approx  & \mathcal{G}_\mathbf{x_q}  + \mathcal{G}'_\mathbf{x_q} \odot (\mathbf{x_n}-\mathbf{x_q}),
\end{aligned}
\end{equation*}
where $\frac{[\cdot]}{[\cdot]}$ is the  element-wise division, {\small $\mathcal{G}'_\mathbf{x_q}= \frac{\partial \mathcal{G}_\mathbf{x_q}}{\partial \mathbf{x_q}}$} denotes the second-order derivative of a task loss with respect to $\mathbf{x_q}$. The above equation can be  represented as:
\begin{equation}
 \label{eq8}
    \mathcal{G}_\mathbf{x_n}  \approx  \mathcal{G}_\mathbf{x_{q}}  \odot\left(1+\frac{\mathcal{G}'_\mathbf{x_q}}{|\mathcal{G}_\mathbf{x_q}|}  \odot \text{sign}(\mathcal{G}_\mathbf{x_q}) \odot(\mathbf{x_{n}}-\mathbf{x_{q}})\right),
\end{equation}
where $|\cdot|$ denotes the absolute value. Comparing Eq. (\ref{eq7}) and Eq. (\ref{eq8}) suggests that we can  connect $\delta$ with $\frac{\mathcal{G}'_\mathbf{x_q}}{|\mathcal{G}_\mathbf{x_q}|}$, but explicitly forming the Hessian matrix $\mathbf{H}$ (containing all $\mathcal{G}'_{\mathbf{x_q}}$) is computationally infeasible in practice. Instead, recent quantized networks approximate the second-order information by the average Hessian Trace~\cite{dong2020hawq} or top Hessian eigenvalues~\cite{dong2019hawq}. In this work, we summarize  the average trace of Hessian and  $\frac{\mathcal{G}'_\mathbf{x_q}}{|\mathcal{G}_\mathbf{x_q}|}$  as scaling factor:
\begin{equation}
\label{eq88}
    \delta = \frac{\text{Tr}(\mathbf{H})/N}{G},
\end{equation}
where $N$ is the number of diagonal elements in $\mathbf{H}$ and $G$ is an average over the absolute values of  gradients, \textsl{i.e.}, $\mathbb{E}[|\mathcal{G}_\mathbf{x_q}|]$.

\begin{algorithm}
    \small
    \DontPrintSemicolon % Some LaTeX compilers require you to use \dontprintsemicolon instead
    \KwIn{A GNN  $f_{gnn}$, bipartite graph $\mathbf{A}$, bit-width $b$, regularizer $\alpha$.}
    
    \KwOut{Model parameters  $\mathbf{\Theta}$ of  $f_{gnn}$;}

    Initialize $\mathbf{\Theta}$ ;\;

 \For{each mini-batch}
{   
\Comment*[l]{Forward pass}
 Compute node embeddings $\mathbf{e}_{u}$ and $\mathbf{e}_{i}$ by Eq. (2);\;
 Normalize  outputs $\hat{\mathbf{e}}_{u} =\frac{\text{clip}(\mathbf{e}_{u}, l, u) -l }{\Delta}$  (same for $\hat{\mathbf{e}}_{i}$);\;
  Quantize values $\bar{\mathbf{e}}_{u} =\text{round}(\hat{\mathbf{e}}_{u})$ (same for $\bar{\mathbf{e}}_{i}$);\;
  Post-scaling   quantized values $\mathbf{q}_{u} =\bar{\mathbf{e}}_{u} \odot\Delta$ (same for $\mathbf{q}_{i}$);\;
  Compute the BPR loss by Eq. (\ref{eq66});\;
  \Comment*[l]{Backward propagation}
  Compute the gradients   $\mathcal{G}_{\bar{\mathbf{e}}_{u}}$ and $\mathcal{G}_{\bar{\mathbf{e}}_{i}}$ via standard SGD;\;
  Adjust the gradients $\mathcal{G}_{\hat{\mathbf{e}}_{u}}$ and $\mathcal{G}_{\hat{\mathbf{e}}_{i}}$ by Eq. (\ref{eq7}):\;
\qquad \quad {\small $\mathcal{G}_{\hat{\mathbf{e}}_{u}}=\mathcal{G}_{\bar{\mathbf{e}}_{u}} \odot \left(1+\delta \cdot \text{sign}(\mathcal{G}_{\bar{\mathbf{e}}_{u}})\odot({\hat{\mathbf{e}}_{u}}-{\bar{\mathbf{e}}_{u}})\right)$},\;
\qquad \quad  {\small $\mathcal{G}_{\hat{\mathbf{e}}_{i}}=\mathcal{G}_{\bar{\mathbf{e}}_{i}} \odot \left(1+\delta \cdot \text{sign}(\mathcal{G}_{\bar{\mathbf{e}}_{i}})\odot({\hat{\mathbf{e}}_{i}}-{\bar{\mathbf{e}}_{i}})\right)$}.\;
  Compute the trace of Hessian by Hutchinson method~\cite{avron2011randomized};\;
  Update  GNN parameters  $\mathbf{\Theta}$ and the scaling factor $\delta$ by Eq. (\ref{eq88});
}
    \Return { $\mathbf{\Theta}$} 
    \caption{HQ-GNN}
    \label{alg2}
\end{algorithm}

    % \begin{table}
    %     \small
    %     \caption{Dataset statistics.}
    %     \label{tab1}
    %     \begin{tabular}{c|ccc}
    %         \hline
    %         Dataset & \#User & \#Items & \#Interactions  \\ \hline
    %         Gowalla    & 29,858    & 40,981     & 1, 027, 370     \\      
    %         Yelp2018   & 31, 668    & 38, 048     & 1, 561, 406      \\      
    %         Amazon-Book   & 52, 643    & 91, 599    & 2, 984, 108     \\   
    %         Alibaba   & 106, 042    & 53, 591    & 907, 407     \\ \hline 
    %     \end{tabular}
    % \end{table}
    % \vspace{-10px}

 We compute the trace of Hessian via Hutchinson's method~\cite{avron2011randomized}
Given a random vector $\mathbf{v}$, whose elements are i.i.d. sampled from a Rademacher distribution such that $\mathbb{E}[\mathbf{v} \mathbf{v}^\top]=\mathbf{I}$. Then, we have:
\begin{equation*}
\begin{aligned}
\text{Tr}(\mathbf{H}) &= \text{Tr}(\mathbf{H} \mathbb{E}[\mathbf{v} \mathbf{v}^\top])=\mathbb{E}[\text{Tr}(\mathbf{H} \mathbf{v} \mathbf{v}^\top)]\\
&= \mathbb{E}[ \mathbf{v}^\top \mathbf{H} \mathbf{v}] \approx \frac{1}{m} \sum_{i=1}^{m} ({\mathbf{v}^{(i)}}^\top \mathbf{H} \mathbf{v}^{(i)}),
\end{aligned}
\end{equation*}
where $\mathbf{I}$ is the identity matrix. The trace of $\mathbf{H}$ can be estimated by  $\mathbb{E}[ \mathbf{v}^\top \mathbf{H} \mathbf{v}]$, where the expectation can be  obtained by drawing  $m$ random vectors. Note that we can first compute $\mathbf{H}\mathbf{v}$, then  $\mathbf{v}^\top \mathbf{H} \mathbf{v}$ is a simple inner product between $\mathbf{v}$ and $\mathbf{H}\mathbf{v}$. Also, we can obtain $\mathbf{H} \mathbf{v}$ efficiently without computing an exact Hessian matrix as follows:
\begin{equation*}
 \label{eq9}
 \frac{\partial (\mathcal{G}^\top_{\mathbf{x_q}} \mathbf{v})}{\partial \mathbf{x_q}} = \frac{\partial \mathcal{G}^\top_{\mathbf{x_q}}}{\partial \mathbf{x_q}}\mathbf{v} + \mathcal{G}^\top_{\mathbf{x_q}} \frac{\partial \mathbf{v}}{\partial \mathbf{x_q}} = \frac{\partial \mathcal{G}^\top_{\mathbf{x_q}}}{\partial \mathbf{x_q}}\mathbf{v} = \mathbf{H}\mathbf{v},
\end{equation*}
where the first equality is the chain rule, while the second is due to the independence of $\mathbf{v}$ and $\mathbf{x_q}$. As such, the cost of Hessian matrix-vector multiply is the same as one gradient back-propagation.

% Algorithm~\ref{alg1} summarizes the Hutchinson's method for efficiently  computing the trace of Hessian.

\subsection{Model Optimization}
\subsubsection{Loss function}
Based on the $b$-bit representations $\mathbf{q}_{u}$ and $\mathbf{q}_{i}$ from Eq. (\ref{eq5}), we can adopt the inner product to estimate the user’s preference towards the target item as: $\hat{y}_{ui} = 	\langle \mathbf{q}_{u}, \mathbf{q}_{i} \rangle $. Also, we  use   Bayesian Personalized Ranking loss to optimize the model~\cite{kang2019candidate}:
\begin{equation}
\label{eq66}
\mathcal{L}_{B P R}(\mathbf{\Theta})=\sum_{\substack{(u, i) \in \mathcal{O}^{+}, (u,j) \notin \mathcal{O}^{+}}} -\ln \sigma\left(\hat{y}_{u i}-\hat{y}_{u j}\right)+\alpha\|\mathbf{\Theta}\|_{F}^{2},
\end{equation}
where $\sigma(\cdot)$ denotes the sigmoid function, $\mathbf{\Theta}$ denotes the model parameters of GNNs, and $\alpha$ controls the $L_2$ regularization strength. Finally, we briefly summarize our HQ-GNN  in Algorithm~\ref{alg2}.

\subsubsection{Complexity}
Compared to vanilla GNN, 
HQ-GNN has an extra time cost to perform gradient adjustments in Eq. (\ref{eq7}). The computation of Hessian Trace  only requires one gradient back-propagation, which is significantly faster than training the GNN encoder itself~\cite{dong2020hawq}. Thus, HQ-GNN has the same training complexity as its GNN encoder. However, during the inference, we can use  integer-only node embeddings (without post-scaling) to generate the top-$k$ candidates, which has both lower memory footprint and faster inference speed compared  to the vanilla GNN.

    \begin{table}[]
\caption{Dataset statistics.}
\label{tab1}
\begin{tabular}{c|cccc}
\toprule[1.2pt]
Dataset       & Gowalla   & Yelp2018  & Amazon-Book & Alibaba \\ \hline
|User|        & 29,858    & 31,668    & 52,643      & 106,042 \\
|Item|        & 40,981    & 38,048    & 91,599      & 53,591  \\
|Interaction| & 1,027,370 & 1,561,406 & 2,984,108   & 907,407 \\ \toprule[1.2pt]
\end{tabular}
\end{table}

\section{Experiments}
\subsection{Experimental Settings}

\begin{table*}[]
\caption{ Performance comparison (bold and underline represent the best  full-precision and 1-bit quantized models).}
\label{tb2}
\scalebox{0.96}{\begin{tabular}{r|cc|cc|cc|cc}
\toprule[1.2pt]
         & \multicolumn{2}{c|}{Gowalla}    & \multicolumn{2}{c|}{Yelp-2018}  & \multicolumn{2}{c|}{Amazon-Book} & \multicolumn{2}{c}{Alibaba} \\  
Methods  & Recall@50         & NDCG@50           & Recall@50         & NDCG@50           & Recall@50          & NDCG@50        & Recall@50          & NDCG@50     \\ \hline
NGCF     & \textbf{0.159} & \textbf{0.130} & \textbf{0.114} & \textbf{0.054} & \textbf{0.092}  & \textbf{0.065} & \textbf{0.071}  & \textbf{0.033} \\
+HashNet & 0.104            & 0.082            & 0.071            & 0.030           & 0.057           & 0.038        & 0.047           & 0.021       \\
+HashGNN & 0.122            & 0.098            & 0.091            & 0.042           & 0.073            & 0.043        & 0.054          & 0.023       \\
+HQ-GNN  & \underline{0.145}         & \underline{0.112}         & \underline{0.101}           & \underline{0.048}           & \underline{0.081}             & \underline{0.054}         & \underline{0.065}         & \underline{0.029}       \\ \hline
LightGCN & \textbf{0.163} & \textbf{0.134} & \textbf{0.118} & \textbf{0.059} & \textbf{0.098}  & \textbf{0.072}  & \textbf{0.076}  & \textbf{0.036}\\
+HashNet & 0.113          & 0.088         & 0.074          & 0.036          & 0.064             & 0.041     & 0.052           & 0.024           \\
+HashGNN & 0.128           & 0.112           & 0.094         & 0.047         & 0.075            & 0.053         & 0.062           & 0.029       \\
+HQ-GNN  & \underline{0.152}               & \underline{0.122}              & \underline{0.108}          & \underline{0.051}             & \underline{0.089}              & \underline{0.062}         & \underline{0.070}             & \underline{0.032}        \\ \toprule[1.2pt]
\end{tabular}}
\end{table*}

\subsubsection{\textbf{Datasets.}} We evaluate our method on four  public datasets~\cite{wang2019neural,he2020lightgcn,Huang2021}: {Gowalla}\footnote{https://snap.stanford.edu/data/loc-gowalla.html}, Yelp-2018\footnote{https://www.yelp.com/dataset}, Amazon-book\footnote{https://jmcauley.ucsd.edu/data/amazon/}, and Alibaba\footnote{https://github.com/huangtinglin/MixGCF/tree/main/data/ali}.  Their statistics are summarized in Table \ref{tab1}.  For each dataset, we randomly select $80\%$ of historical interactions of each user to construct the training set, and treat the remaining as the test set. From the training set, we randomly select $10\%$ of interactions as the validation set to tune the hyper-parameters.

\subsubsection{\textbf{Baselines and Evaluations.}} To verify the effectiveness of HQ-GNN, we mainly compare with graph-based  models:  NGCF~\cite{wang2019neural}, LightGCN~\cite{he2020lightgcn}, HashNet~\cite{cao2017hashnet} and  HashGNN~\cite{tan2020learning}. For HashNet, HashGNN and HQ-GNN, we can choose any GNN encoder to compute the continuous node embeddings in Eq. (\ref{eq2}).  The comparison against other methods (\textsl{e.g.}, factorization machines) is omitted, since most of them are outperformed by LightGCN. We choose the widely-used Recall$@k$ and NDCG$@k$ as the evaluation metrics~\cite{wang2019neural,he2020lightgcn,Huang2021}. We simply set $k = 50$  in all experiments~\cite{tan2020learning}.

\subsubsection{\textbf{Implementation Details.}}  For all baselines, the embedding size of user/item is searched among $\{16,32,64,128\}$. The hyper-parameters (\textsl{e.g.}, batch size, learning rate) of baselines are initialized as  their original settings and are then carefully tuned to achieve the optimal performance. For HQ-GNN, we search $L_2$ regularizer $\alpha$ within $\{10^{-5}, 10^{-4}, 10^{-3}, 10^{-2}, 10^{-1}\}$. In addition, we determine the upper/lower thresholds (Eq. (\ref{eq3})) by exponential moving averages~\cite{jacob2018quantization}, and set the number of bits $b=1$ in Eq. (\ref{eq5}) for fair comparisons with binary hash methods: HashNet~\cite{cao2017hashnet} and HashGNN~\cite{tan2020learning}.

\subsection{Experimental Results}
\subsubsection{\textbf{Overall Performance.}} 
We present a comprehensive performance comparison between full-precision GNNs and quantization-aware GNNs. We summarize the results in terms of Recall$@50$ and NDCG$@50$ for different datasets  in Table \ref{tb2}.  From the table, we have two major  observations: 1) Among all 1-bit GNNs, our proposed HQ-GNN consistently outperforms both HashNet and HashGNN by a large margin on all four datasets. Clearly, this reveals that our HQ-GNNs provide a meaningful gradient adjustments for non-differentiable quantized function. For example, for LightGCN encoder, HQ-GNN has on average 
$15.80\%$ improvement with respect to Recall$@50$ and over $15.63\%$ improvement with respect to NDCG$@50$, comparing to the state-of-the-art HashGNN.  2) It is not surprised that full-precision GNNs perform better than quantization-aware GNNs in all cases. However, quantization-aware GNNs  benefit from both lower memory footprint  and faster inference speed comparing to vanilla GNN. 

In terms of memory  and inference speed, we have observed similar results as those reported in HashNet~\cite{cao2017hashnet} and HashGNN~\cite{tan2020learning}. This is because our HQ-GNN, with $b=1$, inherits all the benefits of HashGNN. For instance, using binarized embeddings (1 bit) can significantly reduce memory usage as compared to using FP32 embeddings. Moreover, the inference speed of our HQ-GNNs is approximately 3.6 times faster than that of full-precision GNNs because the Hamming distance between two binary embeddings can be calculated efficiently~\cite{tan2020learning}. These features make our HQ-GNN more desirable for large-scale retrieval applications in the industry.

\subsubsection{\textbf{Compared to GTE}}

The STE method propagates the same gradient from an output to an input of the discretizer, assuming that the derivative of the discretizer is equal to 1. In contrast, our GSTE method adopts the Hessian to refine the gradients. To evaluate the effectiveness of our GSTE method, we chose LightGCN as the backbone and quantized its embeddings into 1 bit. The performance on different datasets is summarized in Table~\ref{tb3}. From the table, it is clear that our GSTE method performs better than STE for 1-bit quantization, with improvements ranging from $14.7\%$ to $24.5\%$.

Regarding running time, during the training stage, our GSTE method requires computing the trace of Hessian using Hutchinson's method, which is however fast. From Table~\ref{tb3}, we can see that our GSTE method is slightly slower than STE, which is negligible in practice. During inference, both our GSTE and STE methods have the same speed as both use 1-bit quantized embeddings for retrieval, and the trace of Hessian is not needed in the inference stage.

\begin{figure}
	\begin{center}
		\includegraphics[width=8.2cm]{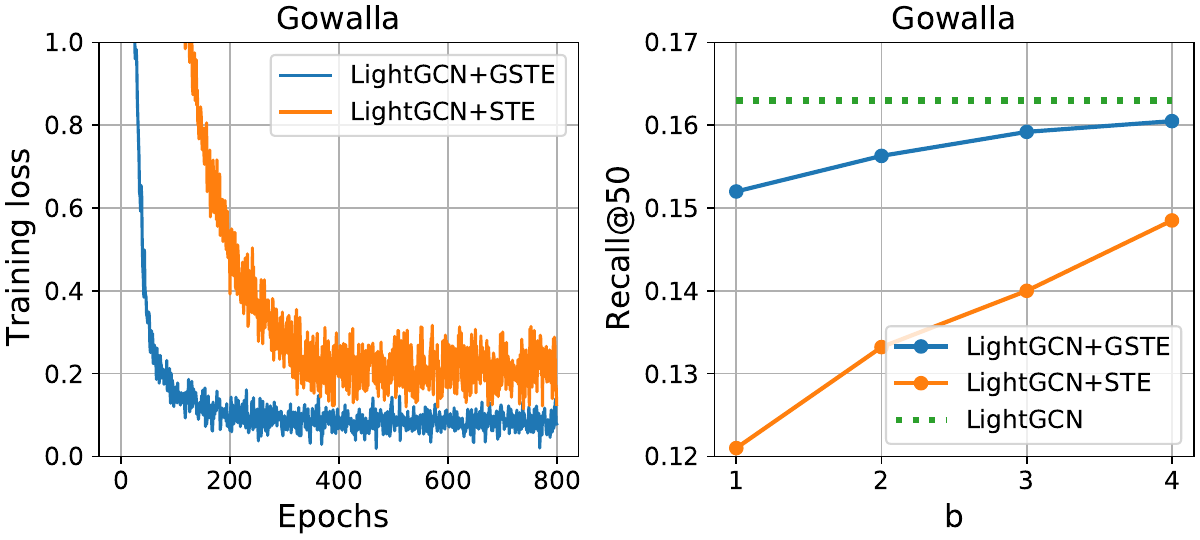}
	\end{center}
	\caption{Left: GSTE vs. STE over training loss. Right: the impact of the number of bits in the HQ-GNN.}
	\label{graph}
\end{figure}

The left of Figure \ref{graph} also displays the training curves of GSTE and STE, and we clearly observe that training quantized LightGCN with GSTE is better than STE in terms of stability. This highlights the effectiveness of utilizing Hessian information in the training process. The right of Figure \ref{graph} shows  the impact of quantization levels by varying $b$ within $\{1,2,3,4\}$ for both GSTE and STE. As can be seen, aggressive quantization (less than 2-bit precision) can lead to significant degradation in the accuracy. When $b=4$, HQ-GNN obtains $98.5\%$ performance recovery of LightGCN. Comparing STE and GSTE, our GSTE consistently performance better than STE in all cases. In summary, HQ-GNN strikes a good balance between latency and performance.

\begin{table*}[]
\caption{  The performance and the running time of 1-bit quantized LightGCN with STE and GSTE.}
\label{tb3}
\scalebox{0.96}{\begin{tabular}{c|cc|cc|cc|cc}
\toprule[1.2pt]
         & \multicolumn{2}{c|}{Gowalla}    & \multicolumn{2}{c|}{Yelp-2018}  & \multicolumn{2}{c|}{Amazon-Book} & \multicolumn{2}{c}{Alibaba} \\  
LightGCN  & Recall@50         & Time(sec)           & Recall@50         & Time(sec)            & Recall@50          & Time(sec)         & Recall@50          & Time(sec)      \\ \hline
+STE    & 0.122 & 30.4 & 0.092 & 41.7 & 0.074  &103.6 & 0.061  & 22.2 \\
+GSTE & 0.152           & 32.9            & 0.108            & 45.1          & 0.089          & 110.7       & 0.070           & 23.9      \\
Improv(\%) & +$24.5\%$          & -           & +$17.3\%$           &-             & +$20.2\%$           &-         & +$14.7\%$          & -      \\\toprule[1.2pt]
\end{tabular}}
\end{table*}

\section{Conclusion}
Training graph neural networks on large-scale user-item bipartite graphs has been a challenging task due to the extensive memory requirement. To address this problem, we propose HQ-GNN that explores the issue of low-bit quantization of graph neural networks for large-scale recommendations. Additionally, we introduce a Generalized Straight-Through Estimator to solve the gradient mismatch problem that arises during the training of quantized networks. HQ-GNN is flexible and can be applied to various graph neural networks. The effectiveness of our proposed method is demonstrated through extensive experiments on real-world datasets.

\bibliographystyle{ACM-Reference-Format}
\balance 
\bibliography{my}
\end{document}